\documentclass[12pt,a4paper,onecolumn]{article}
\usepackage[T1]{fontenc}
\usepackage[left=1in, right=1in, top=1in, bottom=1in]{geometry}
\usepackage{graphicx}
\usepackage{mathtools}
\usepackage{amssymb}
\usepackage{xcolor}
\usepackage{cite}
\usepackage{float}
\usepackage{caption}

\title{Bianchi Type I Space -Time Geometry of the Universe with Time Dependent G and $\Lambda$  Within the Framework of General Relativity: Observational Aspects}

\author{ S. Kotambkar${}^{1}$\footnote{ shubha.kotambkar@gmail.com}, G. K. Goswami${}^{2}$\footnote{ (Corresponding author ) gk.goswami9@gmail.com}, R. Kelkar${}^{3}$\footnote{rupalikelkar@sbjit.edu.in}, G.P. Singh${}^{4}$\footnote{gpsingh@mth.vnit.ac.in },
	\\
	${}^{1}$  Laxminarayan Innovation Technological University,\\ Nagpur, Maharashtra 440033, India \\ 
	${}^{2}$Adjunct Faculty, Department of Mathematics, Madan Mohan Malviya\\
	University of Technology, Gorakhpur (UP),India\\
	${}^{3}$ S. B. Jain Institute of Technology, Management and Research,\\ Nagpur, Maharashtra, India \\
	${}^{4}$ Department of Mathematics, Visvesvaraya National Institute of Technology,\\ Nagpur, Maharashtra, India
}

\begin{document}
	\maketitle
		\begin{abstract}
		Inspired by the latest progress in the hunt of  acceptable cosmological model of the universe, present paper is devoted to explore a mathematical model of the universe having initial anisotropy and with accelerated evolution that attains isotropic character.  We have considered a widely accepted anisotropic Bianchi type I space-time geometry of the universe  to explore a physically viable model. To find an acceptable model, we adopted a hyperbolic form for the scale factor as $ a(t)= \left( sinh bt  \right)^{\frac{1}{n}} $  and a relation $ \Lambda=\Lambda_0 \frac{\ddot{a}}{a}$. The model parameters are constrained by $ \chi^2$ minimization techniques. Using 35 CC measurements we obtain present day expansion rate $H_0= 65.715 Km/S/Mpc$ and the value of $ n= 1.3024$. Further to estimate the local uncertainties of the best fit parameter, we calculate covariance from the Jacobin of the normalized residuals. The corresponding 1 $\sigma$ uncertainties are $ H_0=65.715 \pm 2.392 Km/S/Mpc$ and $n= 1.3024 \pm 0.0842$. The physical and dynamical behaviour of present model has been discussed by a graphical representation of cosmological parameters. The observational constraints on the expansion history are found to be consistent with a present accelerated phase and a transition from deceleration to acceleration at an intermediate redshift. The corresponding evolution of the derived cosmological quantities is also investigated within the adopted model.
	\end{abstract}
	

	\section{Introduction}
The widely trusted space-time for spatially homogeneous and isotropic universe is described as Friedmann-Robertson-Walker (FRW) space-time. The FRW model represents present day universe and it's global approximation.  However the latest observations from different experiments such as cosmic microwave background (CMB) observations and polarization anisotropy foundational \cite{1} cosmic background explores (COBE) \cite{2}, Wilkinson microwave anisotropy investigation \cite{3,4} and plank collaboration \cite{5} provides strong evidence that universe has been anisotropic in the initial phase, which may approach isotropy at late time. This stimulates investigations of anisotropic models as alternative of FRW model. Several authors \cite{6,7} have recommended that anisotropic Bianchi models can play a key role in observational cosmology \cite{8, 9, 10,11,12,13}. By studying Bianchi type models, we can explore the outcomes of anisotropy and dynamics during the cosmic growth. To obtain exact anisotropy information for cosmological investigations researchers prefer Bianchi type I, III, V and IX among the eleven types of anisotropic metric \cite{14}. The majority of the cosmological studies in these metrics are mostly confined to Bianchi type I metric.The Bianchi type I metric may suitably explain the flat spatial geometry of the late time universe. Number of researchers have discussed \cite{15,16,17,18,19,20,21,22,23,24,25,26,27,28} varying G and $\Lambda$ in Bianchi I framework. Very recently Kotambkar et.al. \cite{29,30,31} discussed anisotropic Bianchi I model with G and $\Lambda$  in different context. \\
In Einstein's field equations, the cosmological “constant” $\Lambda$  and the gravitational “constant” G are two fundamental constants that determine how matter and energy interact with the geometry of space-time. Dirac \cite{32} and his large number hypothesis first time proposed idea of a variable G and thereafter it has been considered often in number of modifications of the general theory of relativity. A dynamical $G$ has very interesting consequences in astrophysical and gravitational phenomena \cite{33,34,35,36} . The concept of time dependent $G$ has been comprehensively discussed in the literature \cite{35,36,37,38,39,40,41,42,43}   \\
One of the most tantalizing observational discoveries of the past decade has been that the expansion of the universe is accelerating. This fact was confirmed by different observational surveys such as high redshift supernovae (SNe), cosmic microwave background and baryonic acoustic oscillations to name but a few \cite{44,45,46,47,48,49,50}. The origin of this accelerated expansion was dubbed as “Dark Energy”. Almost 70 \% of the present day energy of the universe consists of dark energy which has large negative pressure. The straightforward model for the dark energy is a cosmological “constant”. Even though the cosmological "constant" has fallen in and out of research interest in the past, but its significance has been studied for the explanation of dark energy. In order to understand accelerating behaviour of the universe, $\Lambda$ played very important role. Taking into consideration the significance of dynamical cosmological "constant" several researchers have discussed different decay law \cite{51,52,53,54,55,56,57}. Abdusattar and Vishwakarma \cite{58} have proposed the conservation of the energy momentum tensor, which as a result offers G and $\Lambda$ as a coupled field, it leaves Einstein’s field equations formally unchanged.

The manuscript is divided into five different sections. In section 1 we write the motivation of present work. The section 2 comprises field equations, Bianchi type I line element with time dependent $G$ and $\Lambda$. In section 3 we discuss cosmological solutions and subsections are devoted to discuss observational constraints, analysis of cosmographic parameters and cosmographic consequences. We summarize the findings in section 4. 

\section{The Cosmological model and Background dynamics}
	The space time geometry of the Bianchi type I metric is given by 

\begin{equation}
	ds^{2}=dt^{2}-R_1^2(t)dx^2-R_2^2(t)dy^2-R_3^2(t)dz^2,
	\label{eq1}
\end{equation}
where directional scale factor $R_1, R_2$ and $R_3$ are the functions of t.
To investigate cosmological behaviour of the universe, Einstein’s field equations with dynamical cosmological and gravitational “constants” are given by
\begin{equation}
	R_i^j-\frac{1}{2} Rg_i^j=-8\pi GT_i^j+\Lambda g_i^j,
	\label{eq2}
\end{equation}
Here $T_i^j$ is the energy momentum tensor.
\begin{equation}
	T_i^j=\left(\rho+p\right)u_iu^j-pg_i^j,
	\label{eq3}
\end{equation}
where p is the perfect fluid pressure and $\rho$ is the matter density , $u_i u^j$ represents fluid four velocity vector such that 
\begin{equation}
	u_iu^j=1.
	\label{eq4}
\end{equation}
In a comoving system of coordinates equation (\ref{eq2}) with metric equation (\ref{eq1}) yields
\begin{equation}
	\frac{\ddot R_1}{R_1}+\frac{\ddot R_2}{R_2}+\frac{\dot R_1 \dot R_2}{R_1 R_2}=-8\pi Gp+\Lambda, 
	\label{eq5}
\end{equation}
\begin{equation}
	\frac{\ddot R_1}{R_1}+\frac{\ddot R_3}{R_3}+\frac{\dot R_1 \dot R_3}{R_1 R_3}=-8\pi Gp+\Lambda, 
	\label{eq6}
\end{equation}
\begin{equation}
	\frac{\ddot R_2}{R_2}+\frac{\ddot R_3}{R_3}+\frac{\dot R_2 \dot R_3}{R_2 R_3}=-8\pi Gp+\Lambda,
	\label{eq7} 
\end{equation}
\begin{equation}
	\frac{\dot R_1 \dot R_2}{R_1 R_2}+	\frac{\dot R_1 \dot R_3}{R_1 R_3}+	\frac{\dot R_2 \dot R_3}{R_2 R_3}= 8\pi G\rho+\Lambda.
	\label{eq8} 
\end{equation}
The dot over a variable represents the derivative with time ‘t’. Differentiation of Eq. (\ref{eq8}), and succeeding simplification of equations (\ref{eq5}) – (\ref{eq7}), leads to

\begin{equation}
	\dot \rho+3\left(\rho+p\right)H+\rho \frac {\dot G}{G}+\frac{\dot \Lambda}{8\pi G}=0.
	\label{eq9}
\end{equation}
The conservation of energy momentum equation of the fluid given by $(T^{;j}_{ij}=0)$ will give
\begin{equation}
	\dot \rho+3\left(\rho+p\right)H=0,
	\label{eq10}
\end{equation}
and thus from equation (\ref{eq9}), we have
\begin{equation}
	8\pi \dot G \rho+\dot \Lambda=0.
	\label{eq11}
\end{equation}
The spatial volume associated with the Bianchi type I universe is given by
\begin{equation}
	V=R_1 R_2 R_3,
	\label{eq12}
\end{equation}
The average scale factor ‘a ’ takes the form
\begin{equation}
	a=\left(R_1 R_2     R_3\right)^\frac{1}{3}=V^\frac{1}{3}.
	\label{eq13}
\end{equation}
The mean Hubble parameter describing the expansion rate of the universe is
\begin{equation}
	H=\frac{1}{3} \left(H_x+H_y+H_z\right),
	\label{eq14}
\end{equation}
Where $ H_i, i=x,y,z$  designate directional Hubble parameter, defined as $ H_i=\frac{\dot{R_i}}{R_i} $ and hence equation (\ref{eq13}) and (\ref{eq14}) gives us
\begin{equation}
	H=\frac{\dot a}{a}=\frac{1}{3}\left(\frac{\dot R_1}{R_1}+\frac{\dot R_2}{R_2}+\frac{\dot R_3}{R_3}\right).
	\label{eq15}
\end{equation}
 The expansion scalar $\Theta$ , which  measures  the rate of expansion or shrinking of the spatial volume, is given by
\begin{equation}
	\Theta=3 H.
	\label{eq16}
\end{equation}
The shear ($ \sigma^2$)  is characterized by a trace-free symmetric tensor. For Bianchi type I the shear is given by


\begin{equation}
\sigma^2=\frac{1}{2} \left[H_x^2+H_y^2+H_z^2\right]-\frac{\Theta^2}{6}.
	\label{eq17}
\end{equation}
The mean anisotropy parameter, which quantifies the deviation of the directional expansion rates from the mean expansion rate, is defined as

\begin{equation}
	A_m=\frac{1}{3}\sum_{i=1}^{3}\left(\frac{H_i-H}{H}\right)^2.
	\label{eq18}
\end{equation}
The deceleration parameter provides a measure of how the expansion rate of the universe develops with time and is defined by
\begin{equation}
	q=-1-\frac{\dot H}{H^2}.
	\label{eq19}
\end{equation}
The generalized Friedmann equation for the present Bianchi type I model is given by
\begin{equation}
	3H^{2}=8\pi G\rho+\Lambda+\sigma^{2},
	\label{eq20}
\end{equation}


 Equation (\ref{eq20}) one may also write as
\begin{equation}
	\rho=\frac{3H^{2}-\Lambda-\sigma^{2}}
	{8\pi G}.
	\label{eq21}
\end{equation}

The conservation equation relating the gravitational constant and the cosmological constant is given by equation (\ref{eq11}). By use of equations ~(\ref{eq11}) and ~(\ref{eq21}), one can get
\begin{equation}
	8\pi\left(\frac{3H^{2}-\Lambda-\sigma^{2}}
	{8\pi G}\right)\dot{G}
	+\dot{\Lambda}=0,
	\nonumber
\end{equation}
or equivalently,
\begin{equation}
	\left(3H^{2}-\Lambda-\sigma^{2}\right)
	\frac{\dot{G}}{G}
	+\dot{\Lambda}=0,
	\nonumber
\end{equation}

Hence,
\begin{equation}
	\frac{\dot{G}}{G}
	=
	-\frac{\dot{\Lambda}}
	{3H^{2}-\Lambda-\sigma^{2}}.
	\label{eq22}
\end{equation}

we have
\[
\dot{G}=\frac{dG}{dz}\dot{z},
\qquad
\dot{\Lambda}=\frac{d\Lambda}{dz}\dot{z}\]

By substituting $\dot{G}$ and $\dot{\Lambda}$ values in equation~(\ref{eq22}), we get
\begin{equation}
	\frac{1}{G}\frac{dG}{dz}
	=
	-\frac{1}
	{3H^{2}-\Lambda-\sigma^{2}}
	\frac{d\Lambda}{dz},
	\label{eq23}
\end{equation}

Integrating both sides, one can get 
\begin{equation}
	\int_{G_0}^{G(z)}\frac{dG}{G}
	=
	-
	\int_{0}^{z}
	\frac{\Lambda'(u)}
	{3H^{2}(u)-\Lambda(u)-\sigma^{2}(u)}
	\,du,
	\label{eq24}
\end{equation}
where
\[
\Lambda'(u)=\frac{d\Lambda}{du}.
\]

We get an expression for gravitational "constant" as
\begin{equation}
	G(z)
	=
	G_0
	\exp\left[
	-
	\int_{0}^{z}
	\frac{\Lambda'(u)}
	{3H^{2}(u)-\Lambda(u)-\sigma^{2}(u)}
	\,du
	\right].
	\label{eq25}
\end{equation}
\section{The Physical and Cosmographic Parameters:}
The Einstein field equations (\ref{eq5})--(\ref{eq8}) constitute a system of four independent equations involving the seven unknown functions
$R_1$, $R_2$, $R_3$, $G$, $\rho$, $p$, and $\Lambda$.
Hence, three additional physically motivated conditions are required to obtain a closed system.
We assume the following hyperbolic form of the average scale factor \cite{58a,59,60,61}, as

\begin{equation}
	a(t)= (\sinh (bt))^{1/n},
	\label{eq26}
\end{equation}

Using the relation
\[
a=\frac{1}{1+z},
\]
where \(z\) is the cosmological redshift.\\

 By use of equations ~(\ref{eq15}), (\ref{eq26}) and the relation $ a=\frac{1}{1+z}$ the Hubble parameter for present model is given by
 \begin{equation}
 	H(z)= \frac{b}{n} \sqrt{(1+z)^{2n}+1}.
 	\label{eq27}
\end{equation}
	
At the present epoch (\(z=0\)), $H_0=\frac{b}{n}\sqrt{2},$ where \(H_0\) denotes the present value of the Hubble parameter. Equation (\ref{eq27}) takes the new form given by

 \begin{equation}
 	H(z)= H_0 \sqrt{\frac{(1+z)^{2n}+1}{2}},
 	\label{eq28}
 \end{equation}
By using equations (\ref{eq19}) and (\ref{eq28}), one can obtain the  deceleration parameter given by 
 \begin{equation}
 	q(z)=-1+ \frac{n (1+z)^{2n}}{(1+z)^{2n}+1}.
 	\label{eq29}
 \end{equation}

  Motivated by several dynamical dark energy models, we assume that the cosmological term is proportional to the cosmic acceleration.
  \[\Lambda \propto \frac{\ddot{a}}{a}\],
  
   Normalizing the relation with respect to its present value, we obtain
  
  \begin{equation}
  	\Lambda(z)=M
  	\left[
  	(1-n)(1+z)^{2n}+1
  	\right],
  	\nonumber
  \end{equation}
  at $z=0$, $\Lambda_0=(2-n)M, M=\frac{b^2}{n^2}. $
  The above equation can be expressed as
  \begin{equation}
  	\Lambda(z)=\frac{\Lambda_0}{2-n}
  	\left[
  	(1-n)(1+z)^{2n}+1
  	\right].
  	\label{eq30}
  \end{equation}
 On combining the spatial Einstein field equations
 (\ref{eq5})--(\ref{eq7}), we obtain
 \begin{equation}
 	\frac{d}{dt}
 	\left(
 	\frac{\dot R_2}{R_2}
 	+
 	\frac{\dot R_3}{R_3}
 	\right)
 	+
 	\left(
 	\frac{\dot R_2}{R_2}
 	+
 	\frac{\dot R_3}{R_3}
 	\right)^2
 	=
 	2\frac{d}{dt}\left(\frac{\dot R_1}{R_1}\right)
 	+
 	2\left(\frac{\dot R_1}{R_1}\right)^2
 	+
 	\frac{\dot R_1}{R_1}
 	\left(
 	\frac{\dot R_2}{R_2}
 	+
 	\frac{\dot R_3}{R_3}
 	\right).
 	\label{eq31}
 \end{equation}
 
 Defining
 \begin{equation*}
 	A=\frac{\dot R_1}{R_1},
 	\qquad
 	B=\frac{\dot R_2}{R_2}+\frac{\dot R_3}{R_3},
 \end{equation*}
   equation (\ref{eq31}) can be rewritten as
 \begin{equation*}
 	\frac{d}{dt}(B-2A)+(A+B)(B-2A)=0,
 \end{equation*}
 For $B-2A\neq0$, this gives
 \begin{equation*}
 	\frac{\dot B-2\dot A}{B-2A}+A+B=0,
 \end{equation*}
 Since
 \begin{equation*}
 	A+B=\frac{d}{dt}\ln(R_1R_2R_3),
 \end{equation*}
 which on integration, yields
 \begin{equation*}
 	(B-2A)R_1R_2R_3=L,
 \end{equation*}
 where $L$ is an integration constant. Hence,
 \begin{equation}
 		\left[
 		\frac{\dot R_2}{R_2}
 		+
 		\frac{\dot R_3}{R_3}
 		-
 		2\frac{\dot R_1}{R_1}
 		\right]
 		R_1R_2R_3=L.
 	\label{eq32}
 \end{equation}
 
 The integration constant $L$ characterizes the deviation from the
 isotropic expansion encoded in the difference between the directional
 Hubble rates. Since the present universe is observationally known to
 be very close to isotropic, we assume that the present day directional
 expansion rates are approximately equal,
 \begin{equation*}
 	H_x\simeq H_y\simeq H_z\simeq H_0.
 \end{equation*}
 Consequently, at the present epoch,
 \begin{equation*}
 	\left[
 	\frac{\dot R_2}{R_2}
 	+\frac{\dot R_3}{R_3}
 	-2\frac{\dot R_1}{R_1}
 	\right]_0\simeq0,
 \end{equation*}
 which motivates the choice
 \begin{equation*}
 	L=0.
 \end{equation*}
 It is important to note that the condition $L=0$ does not imply that
 the Bianchi type-I space-time becomes completely isotropic. The
 anisotropic character of the model is retained through the relative
 evolution of the directional scale factors. 
 
 Equation~(\ref{eq32}) then reduces to
 \begin{equation*}
 	\frac{\dot R_2}{R_2}
 	+\frac{\dot R_3}{R_3}
 	-2\frac{\dot R_1}{R_1}=0,
 \end{equation*}
 on integrating above equation, one can get
 \begin{equation*}
 	R_2R_3=C R_1^2,
 \end{equation*}
 where $C$ is an integration constant. By a suitable rescaling of the
 spatial coordinates, we set $C=1$, yielding
 \begin{equation}
 	R_1^2=R_2R_3.
 	\label{eq33}
 \end{equation}

To satisfy equation (\ref{eq33}), we introduce the parameter

\begin{equation}
	R_2=R_1D, \qquad
	R_3=\frac{R_1}{D},
	\label{eq34}
\end{equation}
where $D = D(t~or~z)$. $D(t)$ measures the deviation from isotropic expansion, with
$D(t)=1$ corresponding to the isotropic limit. 
\\
Using equations~(\ref{eq13}) and (\ref{eq34}), it follows immediately that
\begin{equation}
	R_1=a,\qquad
	R_2=aD,\qquad
	R_3=\frac{a}{D}.
	\label{eq35}
\end{equation}

Using equation ~(\ref{eq35}) and the field equations ~(\ref{eq6}) and (\ref{eq7}), we obtain
\begin{equation}
	\frac{d}{dt}\left(a^{3}\frac{\dot D}{D}\right)=0,
	\label{eq36}
\end{equation}

on integration equation ~(\ref{eq36}), gives
\begin{equation}
	\frac{\dot D}{D}=\frac{K}{a^{3}},
	\label{eq37}
\end{equation}
where \(K\) is an integration constant.

Since \(a=(1+z)^{-1}\), equation ~(\ref{eq37}) becomes
\begin{equation}
	\frac{\dot D}{D}=K(1+z)^3.
	\label{eq38}
\end{equation}

By use of equations (\ref{eq17}) and (\ref{eq35}) the shear scalar takes the simple form
\begin{equation}
	\sigma^{2}=K^{2}(1+z)^{6}.
	\label{eq39}
\end{equation}
The present universe is observed to be nearly isotropic. Accordingly,
we retain a small but non-zero anisotropic contribution through the
constant $K$, choosing
\begin{equation}
	\frac{K}{H_0}\sim10^{-5}.
	\label{eq40}
\end{equation}
Hence, the model preserves the anisotropic Bianchi type I geometry while
allowing the present-day anisotropy to remain sufficiently small,
consistent with the observed near isotropy of the universe.
\\
From equations (\ref{eq25}) and (\ref{eq39}), $G(z)$ can be expressed as
\begin{equation}
	G(z)
	=
	G_0
	\exp\left[
	-
	\int_{0}^{z}
	\frac{\Lambda'(u)}
	{3H^{2}(u)-\Lambda(u)-K^2 (1+u)^6}
	\,du
	\right].
	\label{eq41}
\end{equation}

\subsection{Statistical Method}

Assuming that the observational measurements are statistically
independent and that their uncertainties can be described by Gaussian
errors, we define the chi-square function as
\begin{equation}
	\chi_H^2(H_0,n) = \sum_{i=1}^{35}
	\frac{
		\left[
		H_{\rm obs}(z_i)-H_{\rm th}(z_i;H_0,n)
		\right]^2
	}
	{\sigma_{H_i}^{,2}},
	\label{eq42}
\end{equation}

where $H_{\rm obs}(z_i)$ and $\sigma_{H_i}$ represent the observed Hubble
parameter and its corresponding uncertainty, respectively, while
$H_{\rm th}(z_i;H_0,n)$ is calculated from equation~(\ref{eq28}).
The initial parameter constraints were obtained by minimizing
equation ~(\ref{eq42}). The best-fit values obtained from the 35 cosmic chronometer (CC)
measurements are
\begin{equation}
	H_0=65.715\ {\rm km,s^{-1},Mpc^{-1}},
	\qquad
	n=1.3024.
	\label{eq43}
\end{equation}
The minimum chi-square is
\begin{equation}
	\chi^2_{\min}=15.2297,
	\label{eq44}
\end{equation}
with degrees of freedom
\begin{equation}
	\nu=N-k=35-2=33.
	\label{eq45}
\end{equation}
 By use of equations (\ref{eq44}) and (\ref{eq45}), reduced chi-square is calculated as 
\begin{equation}
	\chi^2_{\rm red}
	=
	\frac{\chi^2_{\min}}{\nu}
	=
	\frac{15.2297}{33}
	=
	0.4615.
	\label{eq46}
\end{equation}
The relatively small value of $\chi^2_{\rm red}$ reflects, in part, the
substantial uncertainties associated with several of the CC measurements.

To estimate the local uncertainties of the best-fit parameters, we also
calculated the covariance matrix from the Jacobian of the normalized
residuals. This gives
\begin{equation}
	{\cal C}=
	\begin{pmatrix}
		5.7235 & -0.1605\\
		-0.1605 & 0.00709
	\end{pmatrix},
	\label{eq47}
\end{equation}
for the parameter vector $(H_0,n)$. The corresponding $1\sigma$
uncertainties are
\begin{equation}
	H_0=65.715\pm2.392
	\ {\rm km,s^{-1},Mpc^{-1}}, \qquad n=1.3024\pm0.0842.
	\label{eq48}
\end{equation}
The correlation coefficient between $H_0$ and $n$ is given by
\begin{equation}
	r_{H_0n}=-0.797,
	\label{eq49}
\end{equation}
Negative sign $r_{H_0n}$ indicates a strong negative correlation between $H_0$ and $n$.
This correlation motivates a full posterior analysis using the Markov
Chain Monte Carlo (MCMC) method.\\
\begin{figure}[H]
	\centering
	\includegraphics[width=0.55\textwidth]{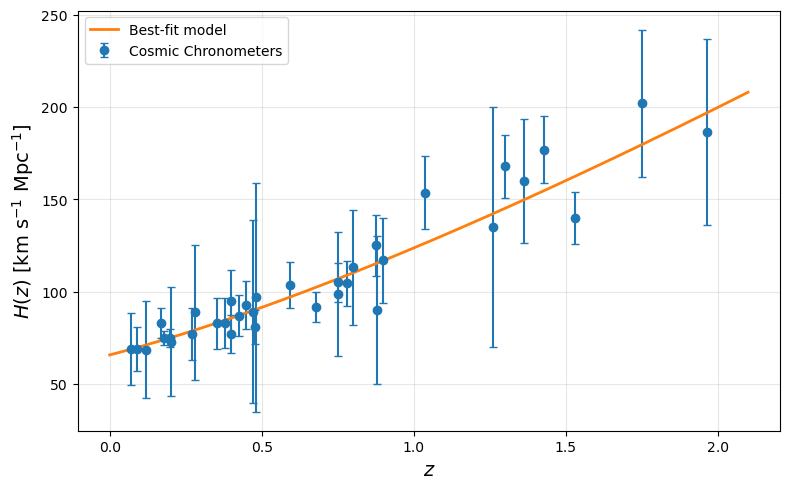}
	\caption{The observational Hubble parameter $H(z)$ from the 35 CC measurements. The error bars represent the reported
		$1\sigma$ uncertainties, while the solid curve represents the
		best-fit theoretical prediction obtained from
		$H(z)=H_0\sqrt{[1+(1+z)^{2n}]/2}$.\\
		The comparison between the
		theoretical Hubble parameter and the observational cosmic chronometer data is depicted in Figure~\ref{fig:Hubble_fit} . The solid curve represents the model prediction obtained from the Hubble function given by equation ~(\ref{eq28}), whereas the data
		points with vertical error bars correspond to the observed $H(z)$ measurements with their associated $1\sigma$ uncertainties,
		as listed in Table~\ref{tab:appendix_hubble}.
	}
	\label{fig:Hubble_fit}
\end{figure}

\subsection{MCMC Analysis}

We further constrain the parameters $(H_0,n)$ using an MCMC analysis.
The posterior probability is written as
\begin{equation}
	{\cal P}(H_0,n|{\cal D})
	\propto
	{\cal L}({\cal D}|H_0,n),
	\Pi(H_0,n),
	\label{eq50}
\end{equation}
where ${\cal D}$ denotes the CC data, ${\cal L}$ is the likelihood and
$\Pi$ represents the prior probability.

For Gaussian observational errors, the likelihood is given by
\begin{equation}
	{\cal L}(H_0,n)
	\propto
	\exp\left[-\frac{1}{2}\chi_H^2(H_0,n)\right].
	\label{eq51}
\end{equation}
We adopt broad uniform priors
\begin{equation}
	40<H_0<90
	\ {\rm km,s^{-1},Mpc^{-1}},
	\qquad
	0.1<n<3.
	\label{eq52}
\end{equation}
The MCMC chains were generated using the affine-invariant ensemble
sampler implemented in the \texttt{emcee} package. We used 50 walkers
and $10,000$ steps per walker. The walkers were initialized in a small
region around the best-fit values obtained from the $\chi^2$ minimization.
The first 2000 steps were discarded as burn-in, and the chains were
subsequently thinned by a factor of 10.

The mean acceptance fraction of the chains is
\begin{equation}
	\langle f_{\rm acc}\rangle=0.7144,
	\label{eq53}
\end{equation}
while the estimated autocorrelation times are
\begin{equation}
	\tau_{H_0}\simeq30.57,
	\qquad
	\tau_n\simeq30.82.
	\label{eq54}
\end{equation}
These values indicate that the adopted chain length is sufficiently
large compared with the characteristic autocorrelation scale. After
burn-in and thinning, a total of $40,000$ posterior samples were used
to estimate the marginalized parameter constraints.

The marginalized posterior distributions give
\begin{equation}
	H_0=
	65.8408^{+2.4093}_{-2.3531}
	\ {\rm km,s^{-1},Mpc^{-1}}, \qquad n=
	1.2972^{+0.0832}_{-0.0866}.
	\label{eq55}
\end{equation}
The close agreement between these constraints and the local covariance estimates confirms the stability of the parameter determination.

\begin{figure}[H]
	\centering
	\includegraphics[width=0.55\textwidth]{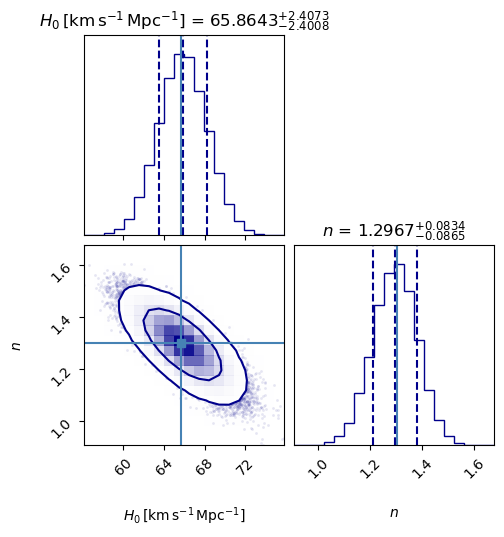}
	\caption{The marginalized posterior distributions and joint confidence
		contours for the parameters $H_0$ and $n$ obtained from the 35 CC measurements. The contours correspond to the $68\%$ and
		$95\%$ confidence regions.}
	\label{fig:corner_H0_n}
\end{figure}

The corresponding two-dimensional posterior distribution in the
$(H_0,n)$ parameter space, together with the marginalized one-dimensional
distributions, is shown in Figure ~\ref{fig:corner_H0_n}. The elongated
orientation of the joint confidence contours reflects the strong
negative correlation between $H_0$ and $n$ found from the covariance
analysis.

\subsection{Cosmographic Consequences}

The deceleration parameter for the present model is given by equation (\ref{eq29}).\\ 
For present epoch, $z=0$, equation.~(\ref{eq29}) reduces to
\begin{equation}
	q_0=-1+\frac{n}{2}.
     \label{eq56}
\end{equation}
Using the MCMC posterior samples of $n$, we obtain
\begin{equation}
	q_0=
	-0.3514^{+0.0416}_{-0.0433}.
     \label{eq57}
\end{equation}
The negative value of $q_0$ indicates that the universe is presently
undergoing accelerated expansion in the present model.

The transition between the decelerating and accelerating phases occurs
when $q(z_t)=0$. From equation ~(\ref{eq29}), this condition gives
\begin{equation}
	z_t=
	\left(\frac{1}{n-1}\right)^{1/(2n)}-1,
\label{eq58}
\end{equation}
Since a real positive transition redshift requires $n>1$, we evaluate
equation ~(\ref{eq58}) for each MCMC sample satisfying this
condition. The resulting posterior distribution yields
\begin{equation}
	z_t=
	0.5961^{+0.3060}_{-0.1769}.
\label{eq59}
\end{equation}
Furthermore, $99.94\%$ of the MCMC posterior samples satisfy $n>1$.
Thus, the CC observations strongly favor the existence of a transition
from an early decelerating phase to the present accelerating phase in
the framework of the present model.\\
For completeness, the parameter $b$ appearing in the original
hyperbolic scale factor is not an independent parameter once $H_0$ and
$n$ have been determined. \\
From
\begin{equation}
	H_0=\frac{b}{n}\sqrt{2},
	\label{eq60}
\end{equation}
we have
\begin{equation}
	b=\frac{nH_0}{\sqrt{2}}.
	\label{eq61}
\end{equation}
Using the best-fit values, we get
\begin{equation}
	b\simeq60.52\ {\rm km,s^{-1},Mpc^{-1}}.
	\label{eq62}
\end{equation}

The observational constraints obtained above provide the first stage of
the validation of the proposed Bianchi type I model. In particular, the
CC data constrain the expansion-history parameters $H_0$ and $n$, which
in turn determine the cosmographic quantities $q(z)$ and $z_t$. These
constraints will subsequently be used in the analysis of the dynamical
cosmological term $\Lambda(z)$ and the varying gravitational constant
$G(z)$.

Figure~\ref{fig:corner_H0_n} presents the marginalized posterior
distributions and joint parameter correlations for the model
parameters $H_0$ and $n$, obtained from the Markov Chain Monte Carlo
(MCMC) analysis of the cosmic chronometer data. The diagonal panels
show the one-dimensional marginalized posterior distributions, while
the off-diagonal panel displays the two-dimensional joint posterior
distribution. The contours correspond to the $68\%$ and $95\%$
credible regions. The posterior distributions yield
\begin{equation}
	H_0=65.84^{+2.41}_{-2.35}
	\ {\rm km\,s^{-1}\,Mpc^{-1}}, \qquad n=1.297^{+0.083}_{-0.087}.
	\label{eq63}
\end{equation}
The two parameters exhibit a negative correlation, indicating that an
increase in $H_0$ is accompanied by a decrease in the preferred value
of $n$. The correlation coefficient obtained from the posterior is
approximately $-0.80$. This correlation reflects the parameter
degeneracy inherent in fitting the Hubble function to the available
cosmic chronometer measurements.
\begin{figure}[H]
	\centering
	\includegraphics[width=0.55\textwidth]{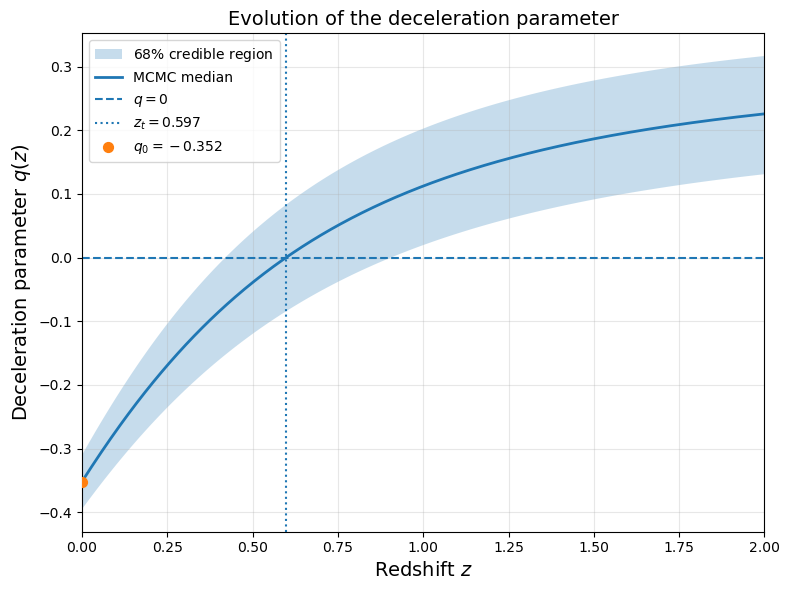}
	\caption{Evolution of the deceleration parameter $q(z)$ as a function
		of redshift for the MCMC median value $n=1.2972$. The horizontal
		dashed line represents $q=0$, while the vertical line indicates the
		transition redshift $z_t=0.5961$. The negative present value
		$q_0=-0.3514$ indicates the current accelerated expansion, whereas
		$q(z)>0$ at higher redshifts corresponds to the earlier decelerating
		phase.}
	\label{fig:qz_transition}
\end{figure}

Figure~\ref{fig:qz_transition} shows the redshift evolution of the
deceleration parameter $q(z)$ obtained by propagating the MCMC
posterior distribution of the parameter $n$. The solid curve denotes
the median prediction, while the shaded region represents the
corresponding $68\%$ credible interval. At sufficiently high redshift,
the deceleration parameter is positive, indicating a decelerated
expansion phase. With the subsequent evolution of the Universe,
$q(z)$ decreases and crosses the $q=0$ line at the transition
redshift
\begin{equation}
	z_t=0.5961^{+0.3060}_{-0.1769}.
	\label{eq64}
\end{equation}
Thus, the model predicts a transition from decelerated expansion
($q>0$) to accelerated expansion ($q<0$) at an intermediate redshift.
At the present epoch, the model gives
\begin{equation}
	q_0=-0.3514^{+0.0416}_{-0.0433},
	\label{eq65}
\end{equation}
which indicates that the universe is currently undergoing accelerated
expansion.

\subsection{Density Parameters and the Evolution of $G$}
\label{subsec:G_evolution}

The observational constraints obtained from the CC data
determine the parameters $H_0$ and $n$. To investigate the subsequent
evolution of the cosmological term and the gravitational constant, we
introduce the present-day density parameters. From the $00$ component of
the Einstein field equations, equation ~(\ref{eq8}), the generalized Friedmann
equation for the Bianchi type I model given by equation (\ref{eq20}).
At the present epoch, equation (\ref{eq20}) may be written as
\begin{equation}
	3H_0^2=8\pi G_0\rho_0+\Lambda_0+\sigma_0^2.
	\label{eq66}
\end{equation}
We define the present day matter, cosmological term and
anisotropy density parameters as
\begin{equation}
	\Omega_{m0}
	=
	\frac{8\pi G_0\rho_0}{3H_0^2},
	\label{eq67}
\end{equation}
\begin{equation}
	\Omega_{\Lambda0}
	=
	\frac{\Lambda_0}{3H_0^2},
	\label{eq68}
\end{equation}
and
\begin{equation}
	\Omega_{\sigma0}
	=
	\frac{\sigma_0^2}{3H_0^2}.
	\label{eq69}
\end{equation}
Consequently the  present day Friedmann equation gives the closure relation as
\begin{equation}
	\Omega_{m0}+\Omega_{\Lambda0}+\Omega_{\sigma0}=1.
	\label{eq70}
\end{equation}
At  present epoch using equations (\ref{eq39}) and (\ref{eq69}), one can obtain
\begin{equation}
	\Omega_{\sigma0}
	=
	\frac{K^2}{3H_0^2}.
	\label{eq71}
\end{equation}
Following the observationally motivated near-isotropy assumption adopted
in our previous analysis, equation (\ref{eq40}) and (\ref{eq71}) yields
\begin{equation}
	\Omega_{\sigma0}
	=
	\frac{1}{3}\left(\frac{K}{H_0}\right)^2
	\simeq3.33\times10^{-11}.
	\label{eq72}
\end{equation}
Equation (\ref{eq72}) indicates that the present anisotropic contribution is negligible.\\

For the present cosmological term, equation (\ref{eq68}), we may express as
\begin{equation}
	\Lambda_0=3H_0^2\Omega_{\Lambda0}.
	\label{eq73}
\end{equation}
For the numerical analysis, we adopt the observationally motivated
value
\begin{equation}
	\Omega_{\Lambda0}=0.70.
	\label{eq74}
\end{equation}
By use of equations (\ref{eq72}) and (\ref{eq74}), equation (\ref{eq70}) yields
\begin{equation}
	\Omega_{m0}\simeq0.30.
	\label{eq75}
\end{equation}
It should be emphasized that these values of $\Omega_{m0}$ and
$\Omega_{\Lambda0}$ are not independently fitted to the 35 CC measurements; rather, $\Omega_{\Lambda0}$ is adopted as a
present-day observationally motivated input, while $\Omega_{m0}$ follows
from the generalized Friedmann constraint.

The normalized Hubble function constrained by the CC
data is
\begin{equation}
	E^2(z)\equiv\frac{H^2(z)}{H_0^2}
	=
	\frac{1+(1+z)^{2n}}{2}.
	\label{eq76}
\end{equation}
The corrected normalized form of the dynamical cosmological term is expressed as
\begin{equation}
	\frac{\Lambda(z)}{3H_0^2}
	=
	\Omega_{\Lambda0}
	\frac{
		1+(1-n)(1+z)^{2n}
	}
	{2-n}.
	\label{eq77}
\end{equation}
On differentiating equation (\ref{eq30}), one can easily obtain
\begin{equation}
	\Lambda'(z)
	=
	\frac{
		2n(1-n)\Lambda_0
	}
	{2-n}
	(1+z)^{2n-1}.
	\label{eq78}
\end{equation}
By use of equations (\ref{eq30}), (\ref{eq76}) and (\ref{eq78}), equation (\ref{eq25}) takes the form
\begin{equation}
	\frac{G(z)}{G_0}
	=
	\exp\left[
	-\int_0^z
	\frac{
		2n(1-n)\Omega_{\Lambda0}(1+u)^{2n-1}/(2-n)
	}
	{
		E^2(u)
		-\Omega_{\Lambda0}
		\left[
		1+(1-n)(1+u)^{2n}
		\right]/(2-n)
		-\left(K/H_0\right)^2(1+u)^6/3
	}
	\,du
	\right].
	\label{eq79}
\end{equation}

The posterior distribution of $n$ obtained from the cosmic chronometer
data can then be propagated into equations.~(\ref{eq77}) and
(\ref{eq79}), allowing the evolution of the cosmological
term and the gravitational constant to be studied together with their
observational uncertainties.
\begin{figure}[htbp]
	\centering
	\includegraphics[width=0.80\textwidth]{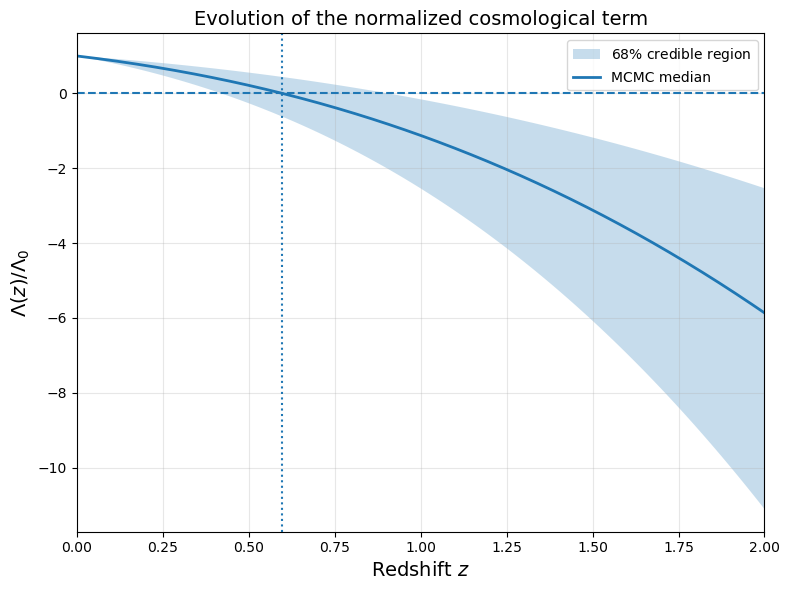}
	\caption{Evolution of the normalized cosmological term
		$\Lambda(z)/\Lambda_0$ obtained from the MCMC posterior of the
		parameter $n$. The shaded region represents the $68\%$ credible
		interval. The horizontal dashed line denotes $\Lambda=0$, while
		the vertical dotted line indicates the transition redshift
		$z_t=0.5961$. The cosmological term changes sign at the same
		redshift at which the universe undergoes the transition from
		decelerated to accelerated expansion.}
	\label{fig:Lambda_evolution}
\end{figure}
\\The evolution of the normalized cosmological term
$\Lambda(z)/\Lambda_0$ is shown in Figure ~\ref{fig:Lambda_evolution}.
The cosmological term decreases from its present positive value and
vanishes at the transition redshift $z_t$. Since the assumed
parametrization satisfies $\Lambda\propto\ddot a/a$, the zero crossing
of $\Lambda(z)$ coincides exactly with the transition from decelerated
to accelerated expansion, i.e.,
\begin{equation}
	\Lambda(z_t)=0
	\quad\Longleftrightarrow\quad
	q(z_t)=0.
	\label{eq80}
\end{equation}
For $z>z_t$, the model predicts $\Lambda(z)<0$, corresponding to the
decelerating phase, whereas $\Lambda(z)>0$ for $z<z_t$, corresponding
to the present accelerating phase.\\
The evolution of the normalized gravitational constant satisfies
$G(0)/G_0=1$. The posterior distribution of $n$ obtained from the
cosmic chronometer analysis is propagated through this equation to
obtain the $68\%$ credible region.

\begin{figure}[htbp]
	\centering
	\includegraphics[width=0.80\textwidth]{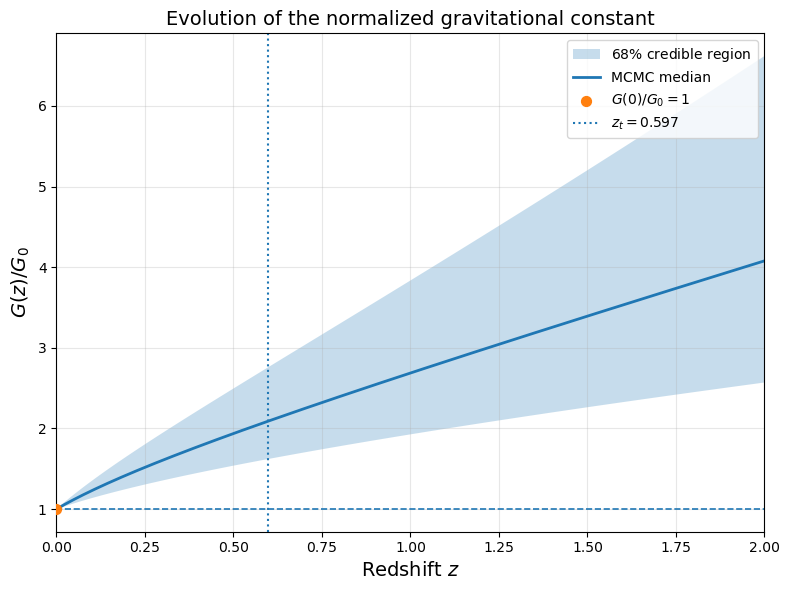}
	\caption{Evolution of the normalized gravitational constant
		$G(z)/G_0$ obtained by propagating the MCMC posterior of the
		parameter $n$ through the generalized conservation equation.
		The shaded region represents the $68\%$ credible interval. The
		horizontal dashed line corresponds to the present value
		$G/G_0=1$, while the vertical dotted line indicates the
		deceleration-to-acceleration transition redshift
		$z_t=0.5961$. The calculation adopts
		$\Omega_{\Lambda0}=0.70$ and $K/H_0=10^{-5}$.}
	\label{fig:G_evolution}
\end{figure}

For the numerical illustration, we adopt the observationally motivated
present day value $\Omega_{\Lambda0}=0.70$ and the near isotropy
condition $K/H_0=10^{-5}$. The resulting evolution of $G(z)/G_0$ is
shown in Figure ~\ref{fig:G_evolution}. The median prediction increases
towards higher redshift, indicating that the gravitational coupling is
larger in the past than at the present epoch within the adopted model.
The increasingly broad credible region at higher redshift reflects the
propagation of the uncertainty in the MCMC-constrained parameter $n$.
The vertical line in Figure ~\ref{fig:G_evolution} marks the transition
redshift $z_t$, at which $\Lambda(z_t)=0$ and $q(z_t)=0$.
It is important to distinguish between the parameters directly
constrained by the cosmic chronometer data and the quantities adopted
for constructing the subsequent evolution of the gravitational
coupling. The parameters $H_0$ and $n$ are directly constrained by the
$H(z)$ observations through the likelihood given in
equation ~(\ref{eq51}). In contrast, the present-day cosmological
density parameter $\Omega_{\Lambda0}$ and the dimensionless anisotropy
parameter $K/H_0$ are not independently constrained by the $H(z)$
dataset considered here. For the purpose of illustrating the
consequences of the model, we adopt the observationally motivated value
$\Omega_{\Lambda0}=0.70$ and the near-isotropy condition
$K/H_0=10^{-5}$. The resulting evolution of $G(z)/G_0$ should therefore
be regarded as an MCMC-propagated model prediction rather than as a
direct observational constraint on the gravitational constant.
Using these inputs together with the MCMC posterior distribution of
$n$, the evolution of the normalized gravitational constant is obtained
from equation ~(\ref{eq79}). The resulting $G(z)/G_0$ is shown
in Figure ~\ref{fig:G_evolution}. The shaded region represents the
propagated $68\%$ credible interval associated with the uncertainty in
$n$. Within the adopted model and assumptions, the median prediction
satisfies
\begin{equation}
	\frac{G(z)}{G_0}>1,\qquad z>0,
	\label{eq81}
\end{equation}
Equation (\ref{eq81}) indicates a larger effective gravitational coupling in the past than
at the present epoch. The uncertainty increases towards higher
redshift because the uncertainty in the MCMC-constrained parameter $n$
is amplified through the nonlinear integral relation for $G(z)$.
The predicted evolution of $G(z)$ is consequently sensitive not only
to the observationally constrained parameter $n$, but also to the
adopted values of $\Omega_{\Lambda0}$ and $K/H_0$. A more complete
observational determination of $G(z)$ would require independent
constraints on these quantities.

\subsection{Evolution of the Matter Density, Pressure, and Equation of State}
\label{subsec:rho_pressure}

The evolution of the matter density and pressure provides important
information about the physical behavior of the cosmological model.
Bianchi type I cosmologies with time-dependent gravitational and
cosmological parameters have been investigated extensively in General
Relativity \cite{31, 62, 63, 64}. In the
present work, we use the observationally constrained background
dynamics to investigate the redshift evolution of the matter density,
pressure, and effective equation-of-state parameter.

The generalized Friedmann equation for the Bianchi type I space-time is given by equation (\ref{eq20}).
It follows directly
from equation ~(\ref{eq20}) that the matter density is given by 
\begin{equation}
	\rho(z)=
	\frac{3H^2(z)-\Lambda(z)-\sigma^2(z)}
	{8\pi G(z)}.
	\label{eq82}
\end{equation}

By use of equations (\ref{eq67}),(\ref{eq76}) and (\ref{eq82}) the normalized matter density can be written as
\begin{equation}
	\frac{\rho(z)}{\rho_0}
	=
	\frac{
		E^2(z)
		-\dfrac{\Lambda(z)}{3H_0^2}
		-\dfrac{\sigma^2(z)}{3H_0^2}
	}
	{
		\Omega_{m0}\dfrac{G(z)}{G_0}
	}.
\label{eq83}
\end{equation}
On simplifying field equations (\ref{eq5})-(\ref{eq7}), one can easily get
\begin{equation}
	2\dot H+3H^2+\sigma^2
	=
	-8\pi Gp+\Lambda.
	\label{eq84}
\end{equation}
We can rewrite equation (\ref{eq19}) as
\begin{equation}
	\dot H=-(1+q)H^2.
	\label{eq85}
\end{equation}
By use of equation ~(\ref{eq85}),equation (\ref{eq84}) can be expressed as
\begin{equation}
	8\pi Gp
	=
	\Lambda+(2q-1)H^2-\sigma^2.
	\label{eq86}
\end{equation}
This relation is particularly useful because it determines the
pressure directly from the background quantities $H(z)$, $q(z)$,
$\Lambda(z)$, and $\sigma^2(z)$, without requiring numerical
differentiation of the reconstructed density.

Using equation ~(\ref{eq67}), the normalized pressure is given by
\begin{equation}
	\frac{p(z)}{\rho_0}
	=
	\frac{
		\dfrac{\Lambda(z)}{3H_0^2}
		+
		\dfrac{1}{3}\left[2q(z)-1\right]E^2(z)
		-
		\dfrac{\sigma^2(z)}{3H_0^2}
	}
	{
		\Omega_{m0}\dfrac{G(z)}{G_0}
	}.
	\label{eq87}
\end{equation}

The corresponding effective equation-of-state parameter is defined by
\begin{equation}
	w(z)=\frac{p(z)}{\rho(z)}.
	\label{eq88}
\end{equation}
Using equations ~(\ref{eq83}) and
(\ref{eq87}), equation (\ref{eq88}) yields
\begin{equation}
	w(z)
	=
	\frac{
		\dfrac{\Lambda(z)}{3H_0^2}
		+
		\dfrac{1}{3}\left[2q(z)-1\right]E^2(z)
		-
		\dfrac{\sigma^2(z)}{3H_0^2}
	}
	{
		E^2(z)
		-\dfrac{\Lambda(z)}{3H_0^2}
		-\dfrac{\sigma^2(z)}{3H_0^2}
	}.
	\label{eq89}
\end{equation}
The pressure is positive near the present epoch and decreases with
increasing redshift. It crosses the zero-pressure line at a relatively
low redshift, approximately $z\simeq0.15$, and subsequently becomes
negative. The effective equation-of-state parameter exhibits a similar
transition, with the median value decreasing from approximately
$w(0)\simeq0.44$ and crossing $w=0$ near $z\simeq0.15$. At higher
redshift, the equation of state approaches approximately
$w(z)\simeq-0.43$ at $z=2$. Thus, the effective fluid associated with
the matter sector exhibits a non-trivial redshift evolution of its
pressure.

It should be emphasized that $\rho(z)$, $p(z)$, and $w(z)$ are derived
quantities rather than independently fitted observational parameters.
The cosmic chronometer data directly constrain the parameters $H_0$
and $n$, while the evolution of $\rho$, $p$, and $w$ follows from the
field equations and the MCMC posterior. In constructing these
quantities, the present-day value $\Omega_{\Lambda0}=0.70$ and the
near-isotropy condition $K/H_0=10^{-5}$ are adopted. Therefore, the
resulting $68\%$ credible regions should be interpreted as propagated
model uncertainties rather than direct observational constraints on
the matter density, pressure, or equation of state.

The behavior of the effective equation of state should also not be
interpreted as that of a pressureless dust component. Instead,
$w(z)$ characterizes the effective perfect-fluid description implied
by the present model and its assumed dynamical $G$ and $\Lambda$.
\begin{figure}[H]
	\centering
	\includegraphics[width=0.80\textwidth]{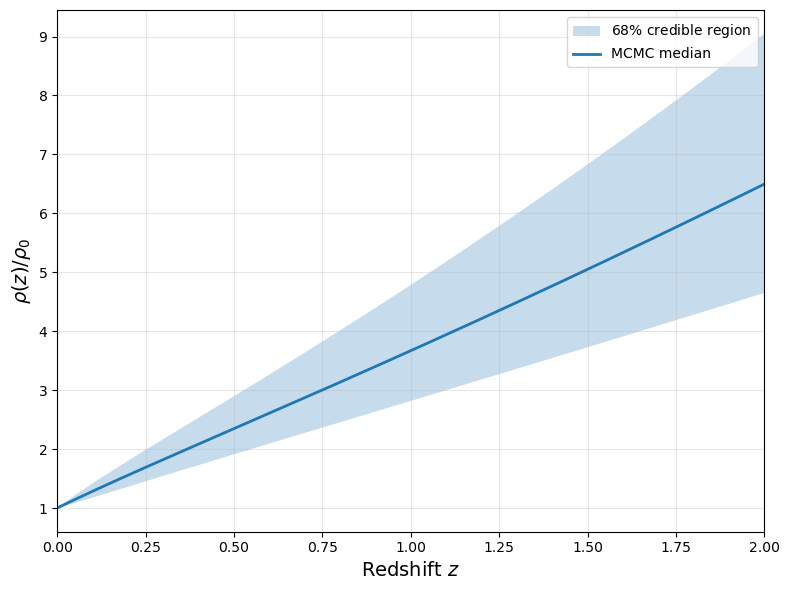}
	\caption{Evolution of the normalized matter density
		$\rho(z)/\rho_0$. The solid curve represents the MCMC median,
		while the shaded region denotes the $68\%$ credible interval
		obtained by propagating the posterior distribution of the
		parameter $n$.}
	\label{fig:rho_evolution}
\end{figure}
\begin{figure}[H]
	\centering
	\includegraphics[width=0.80\textwidth]{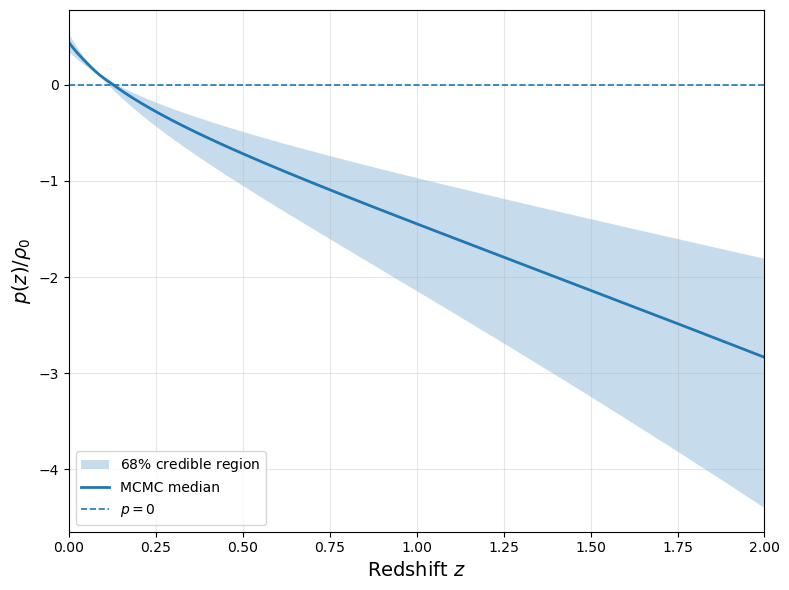}
	\caption{Evolution of the normalized pressure $p(z)/\rho_0$
		obtained directly from the spatial Einstein equations. The solid
		curve denotes the MCMC median and the shaded region represents the
		$68\%$ credible interval. The horizontal dashed line corresponds
		to $p=0$.}
	\label{fig:pressure_evolution}
\end{figure}
\begin{figure}[H]
	\centering
	\includegraphics[width=0.80\textwidth]{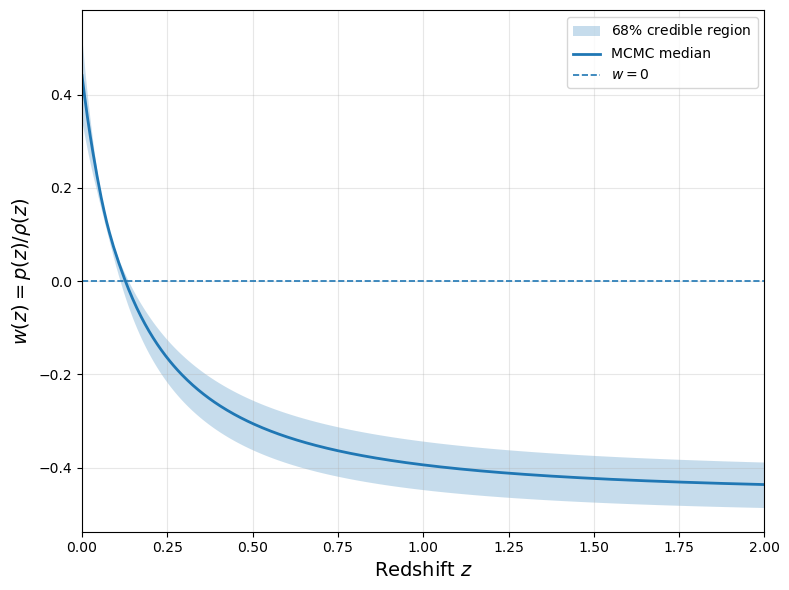}
	\caption{Evolution of the effective equation-of-state parameter
		$w(z)=p(z)/\rho(z)$. The solid curve represents the MCMC median,
		while the shaded region denotes the $68\%$ credible interval.
		The horizontal dashed line indicates $w=0$.}
	\label{fig:w_evolution}
\end{figure}

The evolution of the normalized matter density, pressure, and effective
equation-of-state parameter is presented in
Figures ~\ref{fig:rho_evolution}--\ref{fig:w_evolution}. As shown in
Figure ~\ref{fig:rho_evolution}, the normalized density satisfies
$\rho(0)/\rho_0=1$ and increases monotonically with redshift. This
indicates that the effective matter density was larger in the past.
The uncertainty band becomes broader towards higher redshift, reflecting
the propagation of the uncertainty in the MCMC-constrained parameter
$n$ through the background quantities.

Figure~\ref{fig:pressure_evolution} shows the corresponding evolution
of the normalized pressure. The pressure is positive at the present
epoch and decreases with increasing redshift. It crosses the
zero-pressure line at approximately $z\simeq0.15$ and becomes
increasingly negative at higher redshift. In the present analysis,
the pressure is obtained directly from the spatial Einstein equation (\ref{eq86}), rather than through numerical differentiation of the density. This
provides a direct consistency relation between the pressure and the
background dynamics of the model.

The resulting effective equation-of-state parameter
$w(z)=p(z)/\rho(z)$ is displayed in Figure ~\ref{fig:w_evolution}. The
median curve decreases from approximately $w(0)\simeq0.44$ at the
present epoch, crosses $w=0$ at approximately $z\simeq0.15$, and
approaches $w(z)\simeq-0.43$ at $z=2$. The transition through
$w=0$ therefore occurs at a redshift considerably lower than the
deceleration-to-acceleration transition redshift
$z_t\simeq0.596$. These two transitions have different physical
meanings: the former characterizes the change in the effective
fluid pressure, whereas the latter is defined by $q(z_t)=0$ and
marks the transition in the expansion dynamics.

It is important to emphasize that the quantities
$\rho(z)$, $p(z)$, and $w(z)$ are derived model predictions and are
not independently fitted to the cosmic chronometer data. Their
uncertainties are obtained by propagating the MCMC posterior of the
parameters $H_0$ and $n$, together with the adopted values of
$\Omega_{\Lambda0}$ and $K/H_0$.
\\The jerk j gives rate of change of cosmic acceleration is defined as  $j=\frac{1}{aH^3} \frac{d^3a}{dt^3}$  and the snap parameter (s) is defined as  $s=\frac{1}{aH^4} \frac{d^4a}{dt^4}$ .
\\
For the present model the jerk and the snap parameters are obtained as 
\begin{equation}
	j=\frac{n(3-2n)}{(1+z)^{2n} + 1} +(1-n)(1-2n),
	\label{eq90}
\end{equation}
\begin{equation}
	s=\frac{n^2 (3-2n)}{((1+z)^{2n} + 1)^2} +\frac{2n (1-n)(3-4n)}{(1+z)^{2n} + 1}+(1-n)(1-2n)(1-3n).
	\label{eq91}
\end{equation}
The jerk and snap parameters expressed in terms of redshift characterize the higher-order behaviour of the cosmic expansion.

\section{Conclusion}
In present paper, we have explored Bianchi type I cosmological model with dynamical G and $\Lambda$.  Using the scale factor of the form $ a(t)=(sinh bt)^\frac{1}{n}$ and $\Lambda \propto \frac{\ddot{a}}{a}$. The mean Hubble parameter of the model have been tested with the observational data. To explore the observational viability of the Hubble parameter we use cosmic chronometer data. For the present model best fit values of Hubble parameter  obtained are $H_0= 65.715 Km/Sec/Mpc$ and $n= 1.3024$. To estimate the local uncertainties of the best fit parameters we also calculated 1 $\sigma$ uncertainties, which are $H_0= 65.715 \pm 2.392 Km, S^{-1}, Mpc^{-1}$ and $ n= 1.3024 \pm 0.0842$.The correlation coefficient between $H_0$ and $n$ has been obtained as $ r_{H_0n}=-0.797$, which indicate strong negative correlation between these two parameters. We constrain the parameters using MCMC analysis, the values of $H_0$ and $n$ are, $H_0=65.8408^{+2.4093}_{-2.3531}\ {\rm km,s^{-1},Mpc^{-1}},	n = 1.2972^{+0.0832}_{-0.0866}$. The close agreement between these constraints and the local covariant estimates confirms the stability of the parameter determination. \\
The deceleration parameter at the present epoch, obtained from the MCMC posterior samples of $n$, is
$q_0=-0.3514^{+0.0416}_{-0.0433}$. The negative value indicates that the Universe is presently undergoing accelerated expansion in the proposed model. In this model we got $n= 1.3024$, and $99.94\%$ of the MCMC posterior samples satisfy $ n > 1$. Thus the CC observations strongly support the existence of transition from an early decelerating phase to present accelerating phase for the present model.\\
From Figure \ref{fig:Lambda_evolution}, one observes that $\Lambda(z)$ decreases with increasing redshift. It remains positive for $z<z_t$, vanishes at $z=z_t$, and becomes negative for $z>z_t$. Thus, the sign change of the cosmological term occurs at the same redshift as the
deceleration-to-acceleration transition.\\
Figure \ref{fig:G_evolution} indicates that $\frac{G(0)}{G_0}=1 $. Equation (\ref{eq81}) indicates a larger effective gravitational coupling in the past than at the present epoch within the adopted model.\\
From Figure~\ref{fig:rho_evolution}, one finds that $\rho(0)/\rho_0=1$ and that the matter density increases monotonically with redshift, indicating
that the effective matter density was larger in the past.\\
Figure \ref{fig:pressure_evolution} depicts that the pressure is positive near the present epoch and decreases with increasing redshift which represents the redshift evolution predicted by the present model. At the present epoch, the anisotropic contribution is extremely small, with $\Omega_{\sigma0}\simeq3.33\times10^{-11}$, indicating
that the Universe is very close to isotropy.\\
For flat standard $\Lambda$CDM  $j_0 \approx 1$ and $ s_0 \approx -0.42$, for present model we obtained $j_0=0.74264, s_0=-0.37304$ which closely matches with $\Lambda$CDM model.


\appendix
	\renewcommand{\thefigure}{A\arabic{figure}}
\renewcommand{\thetable}{A\arabic{table}}
\setcounter{figure}{0}
\setcounter{table}{0}
\section*{Appendix A. Hubble Table:}

	\section*{}

The cosmic chronometer $H(z)$ dataset used in this analysis is listed in Table~\ref{tab:appendix_hubble}. The measurements are compiled from Refs.~\cite{Zhang2014,Simon2005,Moresco2012,Moresco2016,Moresco2015,Ratsimbazafy2017,Stern2010,Borghi2022,Jimenez2023,Jiao2023,Tomasetti2023}.

\begin{table}[htbp]
	\centering
	\captionsetup{skip=10pt}
	\caption{Hubble Parameter Measurements from Cosmic Chronometers}
	\label{tab:appendix_hubble}
	
	\begin{tabular}{|c|c|c|c|}
		\hline
		Redshift ($z$) & $H(z)$ [km/s/Mpc] & Error [km/s/Mpc] & Reference \\
		\hline
		0.07 & 69.0 & 19.6 & \cite{Zhang2014} \\
	0.09 & 69.0 & 11.9991 & \cite{Simon2005} \\
	0.12 & 68.6 & 26.2 & \cite{Zhang2014} \\
	0.17 & 83.0 & 8.00037 & \cite{Simon2005} \\
	0.1791 & 74.91 & 3.8069 & \cite{Moresco2012} \\
	0.1993 & 74.96 & 4.9001 & \cite{Moresco2012} \\
	0.2 & 72.9 & 29.6 & \cite{Zhang2014} \\
	0.27 & 77.0 & 13.9986 & \cite{Simon2005} \\
	0.28 & 88.8 & 36.6 & \cite{Zhang2014} \\
	0.3519 & 82.78 & 13.9484 & \cite{Moresco2012} \\
	0.3802 & 83.0 & 13.54 & \cite{Moresco2016} \\
	0.4 & 95.0 & 16.9955 & \cite{Simon2005} \\
	0.4004 & 76.97 & 10.18 & \cite{Moresco2016} \\
	0.4247 & 87.08 & 11.24 & \cite{Moresco2016} \\
	0.4497 & 92.78 & 12.9 & \cite{Moresco2016} \\
	0.4783 & 80.91 & 9.044 & \cite{Moresco2016} \\
	0.47 & 89.0 & 49.6 & \cite{Ratsimbazafy2017} \\
	0.48 & 97.0 & 62.0024 & \cite{Stern2010} \\
	0.5929 & 103.8 & 12.4975 & \cite{Moresco2012} \\
	0.6797 & 91.6 & 7.9619 & \cite{Moresco2012} \\
	0.75 & 98.8 & 33.6 & \cite{Borghi2022} \\
	0.75 & 105.0 & 10.76 & \cite{Jimenez2023} \\
	0.7812 & 104.5 & 12.1951 & \cite{Moresco2012} \\
	0.8 & 113.1 & 31.2 & \cite{Jiao2023} \\
	0.8754 & 125.1 & 16.7009 & \cite{Moresco2012} \\
	0.88 & 90.0 & 39.996 & \cite{Stern2010} \\
	0.9 & 117.0 & 23.0022 & \cite{Simon2005} \\
	1.037 & 153.7 & 19.6736 & \cite{Moresco2012} \\
	1.26 & 135.0 & 65.0 & \cite{Tomasetti2023} \\
	1.3 & 168.0 & 17.0016 & \cite{Simon2005} \\
	1.363 & 160.0 & 33.58 & \cite{Moresco2015} \\
	1.43 & 177.0 & 18.0009 & \cite{Simon2005} \\
	1.53 & 140.0 & 14.0 & \cite{Simon2005} \\
	1.75 & 202.0 & 39.996 & \cite{Simon2005} \\
	1.965 & 186.5 & 50.43 & \cite{Moresco2015} \\
	\hline\hline
	\end{tabular}
	\end{table}
\begin{figure}[H]
	\centering
	\includegraphics[width=0.85\textwidth]{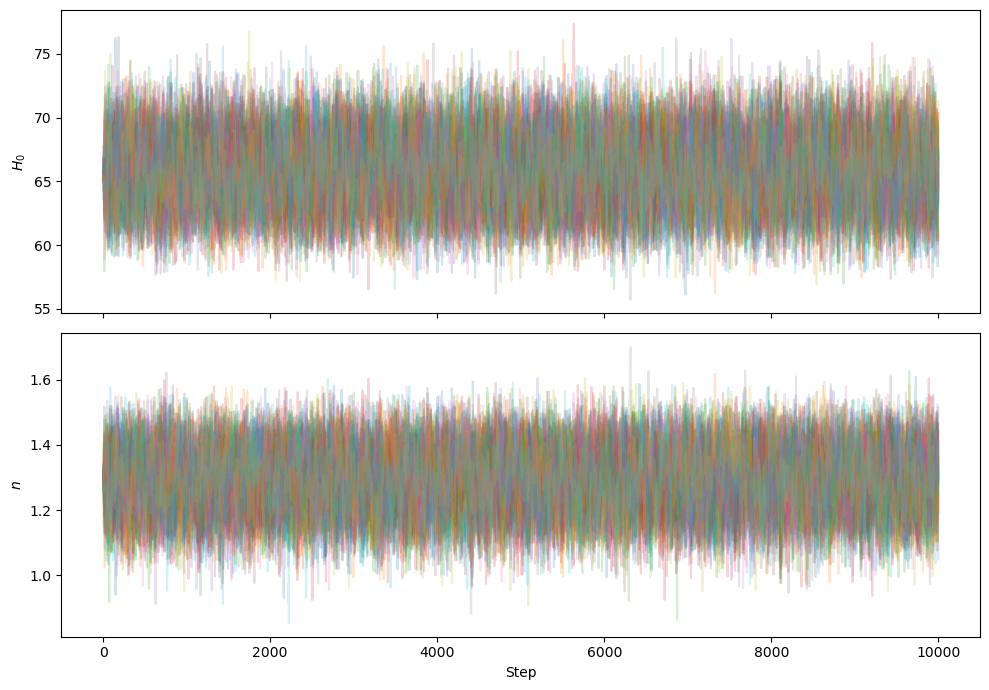}
	\caption{Trace plots of the MCMC chains for $H_0$ and $n$. The chains
		show good mixing and convergence after the initial burn-in phase.}
	\label{fig.A1}
\end{figure}

\end{document}